\documentclass{ws-ijseke}

\usepackage[T1]{fontenc}	
\usepackage[utf8]{inputenc}	
\usepackage{graphicx}	
\usepackage[misc]{ifsym}	
\usepackage{tabularx}	
\usepackage{multirow}	
\usepackage{makecell}
\usepackage[multiple]{footmisc}
\usepackage{amssymb}
\usepackage{ragged2e}
\usepackage[section]{placeins}

\usepackage{url}
\usepackage[sort,compress]{cite}

\begin{document}


%
\catchline{01}{01}{2003}{}{}
%

\title{Automatic Analysis of Available Source Code of Top Artificial Intelligence Conference Papers}

\author{Jialiang Lin\textsuperscript{[1,2]}, Yingmin Wang\textsuperscript{[1,2]}, Yao Yu\textsuperscript{[1]}, Yu Zhou\textsuperscript{[1]}, \\ Yidong Chen\textsuperscript{[1,2]}, Xiaodong Shi\textsuperscript{[1,2]}* \\
\email{me@linjialiang.net, *mandel@xmu.edu.cn, *Corresponding author}
}

\address{[1]School of Informatics, Xiamen University, Xiamen, P. R. China}
\address{[2]Key Laboratory of Digital Protection and Intelligent Processing of Intangible Cultural Heritage of Fujian and Taiwan, Ministry of Culture and Tourism, P. R. China}

\maketitle

\begin{abstract}
Source code is essential for researchers to reproduce the methods and replicate the results of artificial intelligence (AI) papers. Some organizations and researchers manually collect AI papers with available source code to contribute to the AI community. However, manual collection is a labor-intensive and time-consuming task. To address this issue, we propose a method to automatically identify papers with available source code and extract their source code repository URLs. With this method, we find that 20.5\% of regular papers of 10 top AI conferences published from 2010 to 2019 are identified as papers with available source code and that 8.1\% of these source code repositories are no longer accessible. We also create the XMU NLP Lab README Dataset, the largest dataset of labeled README files for source code document research. Through this dataset, we have discovered that quite a few README files have no installation instructions or usage tutorials provided. Further, a large-scale comprehensive statistical analysis is made for a general picture of the source code of AI conference papers. The proposed solution can also go beyond AI conference papers to analyze other scientific papers from both journals and conferences to shed light on more domains.
\end{abstract}


\section{Introduction}

The past decade has witnessed enormous advances in artificial intelligence (AI), which brings along tremendous shifts in all aspects of human life. Face recognition \cite{taskiran-face-2020}, intelligent personal assistants \cite{silva-intelligent-2020} and recommender systems \cite{zhang-deep-2019}, to name a few, are just daily life examples of AI implementation. Today's AI products such as IBM Watson \cite{ferrucci-introduction-2012}, AlphaGo of DeepMind \cite{silver-mastering-2017}, and Pluribus developed jointly by CMU and Facebook \cite{brown-superhuman-2019} can even triumph over the best human players.

In this day and age, knowledge spreads at an incredible speed. It takes only 8 years for a newly proposed concept to be widely accepted by the computer science (CS) community \cite{detecting-lin-2022}. Meanwhile, research in CS, especially in AI, is highly vulnerable to publication delays. The publishing process of most journals is time-consuming \cite{bilalli-framework-2021}, which might negatively affect the novelty of the research. On account of this, numerous researchers choose conferences to publish their papers. A great number of AI technologies, like Adam \cite{kingma-adam-2015}, ResNet \cite{he-deep-2016}, Transformer \cite{vaswani-attention-2017}, ULMFiT \cite{howard-universal-2018}, and BERT \cite{devlin-bert-2019} are published as top conference papers. Research also shows that top conferences are given more priority over top journals in the field of CS \cite{vrettas-conferences-2015}. The last decade has seen a surging trend in the submissions to top AI conferences. For instance, 2010 registered 1,219 submissions to the prestigious Annual Conference on Neural Information Processing Systems (NeurIPS) and in just a decade, the number increased considerably to 6,743 in 2019.\footnote{https://www.openresearch.org/wiki/NIPS}

It is critical for AI papers to be accompanied by available source code to enable the reproduction of the methods and replication of the results because AI research relies heavily on experiments. Unlike theoretical research that can be proved through reasoning, AI research can only be tested by rerunning the experiments with source code, which is preferably released by the authors. Apart from this, numerous factors, including deployment environments, hyperparameters, random seeds, etc., can undermine the reproduction of AI experiments. In recent years, AI researchers are grappling with a reproducibility crisis \cite{hutson-artificial-2018,raff-step-2019}, in which many AI studies are found difficult to be reproduced or replicated. As a result of this crisis, reproducibility becomes a hallmark of high-quality AI papers. Most top AI conferences require authors to provide adequate technical details of reproducibility in their papers. It is also recommended by these conferences to upload corresponding source code when submitting the papers.\footnote{https://neurips.cc/public/guides/CodeSubmissionPolicy}  Lin et al. \cite{lin-how-2020} reveals that providing source code is a distinct feature of published papers compared with unpublished ones in the CS category of arXiv \cite{ginsparg-winners-1997}.

When researchers release their source code, they help readers to reproduce their experiments. More importantly, the contributed source code can serve as a catalytic foundation for future research. That being said, collecting and analyzing AI conference papers with available source code is of great benefit as this work can expand access to more papers of this kind. The collection of available source code published in top AI conference papers has been conducted manually\footnote{Collection through automatic extraction and merging with manually collected data is still considered as manual collection of source code.} by some organizations, e.g. Papers With Code\footnote{https://paperswithcode.com/} and researchers, e.g. Zaur Fataliyev.\footnote{https://github.com/zziz/pwc} However, with the vigorous growth rate of AI conference papers, the manual way of paper collecting can be highly demanding in labor and time. To our best knowledge, there is no automatic solution to identify and collect available source code of AI conference papers and analyze their characteristics. In view of the current need, we raise three research questions (RQs) as follows:

\begin{itemize}

\item RQ1 What percentage of released source code do top AI conference papers have?

\item RQ2 What are the characteristics of these source code?

\item RQ3 Do the authors provide enough documentation for readers to run the source code?

\end{itemize}

By answering these RQs, we make the following contributions:

\begin{itemize}

\item We introduce an automatic method to identify papers with available source code and extract the URLs of their source code repositories from these papers in PDF format. The collected URLs are publicly available;
\item We construct and release the XMU NLP Lab README Dataset. This is the largest README dataset consisting of 5k manually labeled README files on the GitHub repositories of papers with available source code. The labeling tool is also released;
\item We perform the first large-scale analysis of available source code of top AI conference papers over the last decade;

\end{itemize}

A step forward from the work above is to answer the question that whether the results of papers can be replicated with source code in practice. Manually rerunning the source code for every repository is a laboring and time-consuming task \cite{bonneel-code-2020}. Reaching an automatic solution, however, is also not an easy task. It requires the collaboration of many up-to-date technologies, e.g. information extraction and auto-deployment. Thus, this task is still pending for future study.

The rest of this paper is structured as follows. Sections~\ref{sec:rq1}, \ref{sec:rq2}, and~\ref{sec:rq3} answer RQ1, RQ2, and RQ3, respectively. Section~\ref{sec:whole-pipeline} elaborates the pipeline of the whole analysis. Section~\ref{sec:related-work} presents the related work. Last but not least, Section~\ref{sec:future-conclusion} describes our future work and conclusion.

\section{RQ1: What Percentage of Released Source Code Do Top AI Conference Papers Have?}
\label{sec:rq1}

There is an appreciable amount of AI conferences and quite a few of them are considered top conferences. The selection of top AI conferences in our research is not exhaustive. We choose our research objects from the Artificial Intelligence category of the 5th edition of the List of International Academic Conferences and Periodicals Recommended by China Computer Federation (CCF)\footnote{https://www.ccf.org.cn/en/Bulletin/2019-05-13/663884.shtml} and the Artificial Intelligence subcategory of Top Publication of Google Scholar.\footnote{https://scholar.google.com/citations?view\_op=top\_venues\&hl=en\&vq=eng\_artificialintelligence} In total, 10 conferences are selected: the AAAI Conference on Artificial Intelligence (AAAI), the Annual Meeting of the Association for Computational Linguistics (ACL), the IEEE Conference on Computer Vision and Pattern Recognition (CVPR), the European Conference on Computer Vision (ECCV), the Conference on Empirical Methods in Natural Language Processing (EMNLP), the International Conference on Computer Vision (ICCV), the International Conference on Learning Representations (ICLR), the International Conference on Machine Learning (ICML), the International Joint Conference on Artificial Intelligence (IJCAI), and the Annual Conference on Neural Information Processing Systems (NeurIPS). The selected conferences can be divided into three categories depending on their focusing topics: computer vision (CV) conferences (CVPR, ECCV, ICCV), natural language processing (NLP) conferences (ACL, EMNLP), and general AI conferences (AAAI, ICLR, IJCAI, NeurIPS).

The period we study is from 2010 to 2019. The reason why we do not set the period to the year when writing this paper is that some data of our studies take time to accumulate. These data cannot reflect differences in a short term. All the conferences started before 2010, except for the 2013-founded ICLR. Among them, ECCV and ICCV are held biennially. IJCAI takes place every two years before 2015 but every year since then. The remaining seven of these top AI conferences are organized annually. We study regular papers published in these venues and altogether 36,679 papers are collected.

\subsection{Method}
\label{subsec:method}

Suppose a paper $p$ has a URL set $U$. $u_i$ denotes a URL in $U$, $u_i \in U=\{u_1,u_2,\ldots,u_n\} (n > 0)$, where $n$ is the total number of URLs in $U$. When $n = 0$, $U$ is $\varnothing$. A URL $u_{i}$ may link to different resources, namely the paper's source code repository, another source code repository, a cited web page, a tool the paper uses, etc. If any URL $u_{i}$ can be identified as the link of the own source code repository of $p$, then this paper $p$ can be recognized as a paper with available source code. Otherwise, it is not. This problem can be solved by text classification:

\begin{equation}
F = Classification(p)
\end{equation}

That is

\begin{equation}
Classification(p) = \bigcup_{i=1}^{n}URLClassification(u_{i})
\end{equation}

However, a URL $u_{i}$ alone may not have enough information to help identify whether it is the paper's own source code repository or not. On observation of a large quantity of AI papers, we find that the sentence $s_{i}$ containing URL $u_{i}$ will have this kind of information. If the URL is the link of a paper's own source code repository, the sentence that contains this URL will begin with content like "the code is public on...", "we release the implementation and models at...", "the sources of our methods are available from...", etc. If the URL is not the link of this paper's own source code repository, the beginning will be like "those data are available from...", "all proofs can be found in the on-line appendix...", "we thank the authors for releasing their code at...", etc. These distinguishable patterns enable the identification of papers with available source code as we transfer our classification function as follows:

\begin{equation}
Classification(p) = \bigcup_{i=1}^{n}SentenceClassification(s_{i})
\end{equation}

In general, supervised classification models have higher accuracy and trustworthiness than those unsupervised ones. In order to use supervised models, labeled data are required and they are obtained from Papers With Code. Papers With Code links academic papers in the AI field to their corresponding available source code. It provides its daily-updated metadata (paper titles, paper PDF URLs, available source code repository URLs, etc.) in JSON format. In its JSON file,\footnote{The file was downloaded on Jun 4, 2020.} 13,625 of the records are marked as "mentioned\_in\_paper" and this marking indicates that these records have their own available source code repository URLs included in the papers. We download the PDF files of these papers and convert them into XML files using GROBID \cite{lopez-grobid-2009}. We then parse the XML files to locate the source code repository URLs. All the sentences that refer to these URLs are collected as positive samples. Sentences that mention other URLs are randomly collected as negative samples. The negative samples are three times as many as the positive samples. Together the positive and negative samples are randomly split into a training set, a validation set, and a test set at a ratio of 8:1:1.

Out of various models for text classification, we choose Bidirectional Encoder Representations from Transformers (BERT). BERT is an up-to-date pre-trained language model which has been proved to achieve excellent performance on text classification \cite{sun-how-2019}. SciBERT \cite{beltagy-scibert-2019} is a domain-specific BERT version trained on 1.14 million CS and biomedical papers. The authors of SciBERT demonstrate that using SciBERT can achieve better performance than BERT on the scientific text. Hence, SciBERT, optimized for the scientific domain, is finally used.

We mostly follow the same architecture used in Beltagy et al. \cite{beltagy-scibert-2019} for fine-tuning using the training set. Our fine-tuned model achieves an accuracy of 0.939 and an F1-score of 0.909 on our test set. We then download the PDF files of regular papers of the 10 top AI conferences. We follow the same steps to parse these PDF files, use a regular expression to locate all valid URLs starting with "http", "https" or "ftp" in full texts except references, and extract sentences containing these URLs. The rest of the words in these sentences after removing URLs are then fed into our fine-tuned classification model. Based on the classification result, we can judge whether the paper provides available source code or not. Some PDF files of paper cannot be parsed by GROBID. For these papers, we manually check whether they provide available source code and collect the repository URLs if positive.

\subsection{Result}
\begin{figure}[htb]
\centering
\includegraphics[width=0.9\textwidth]{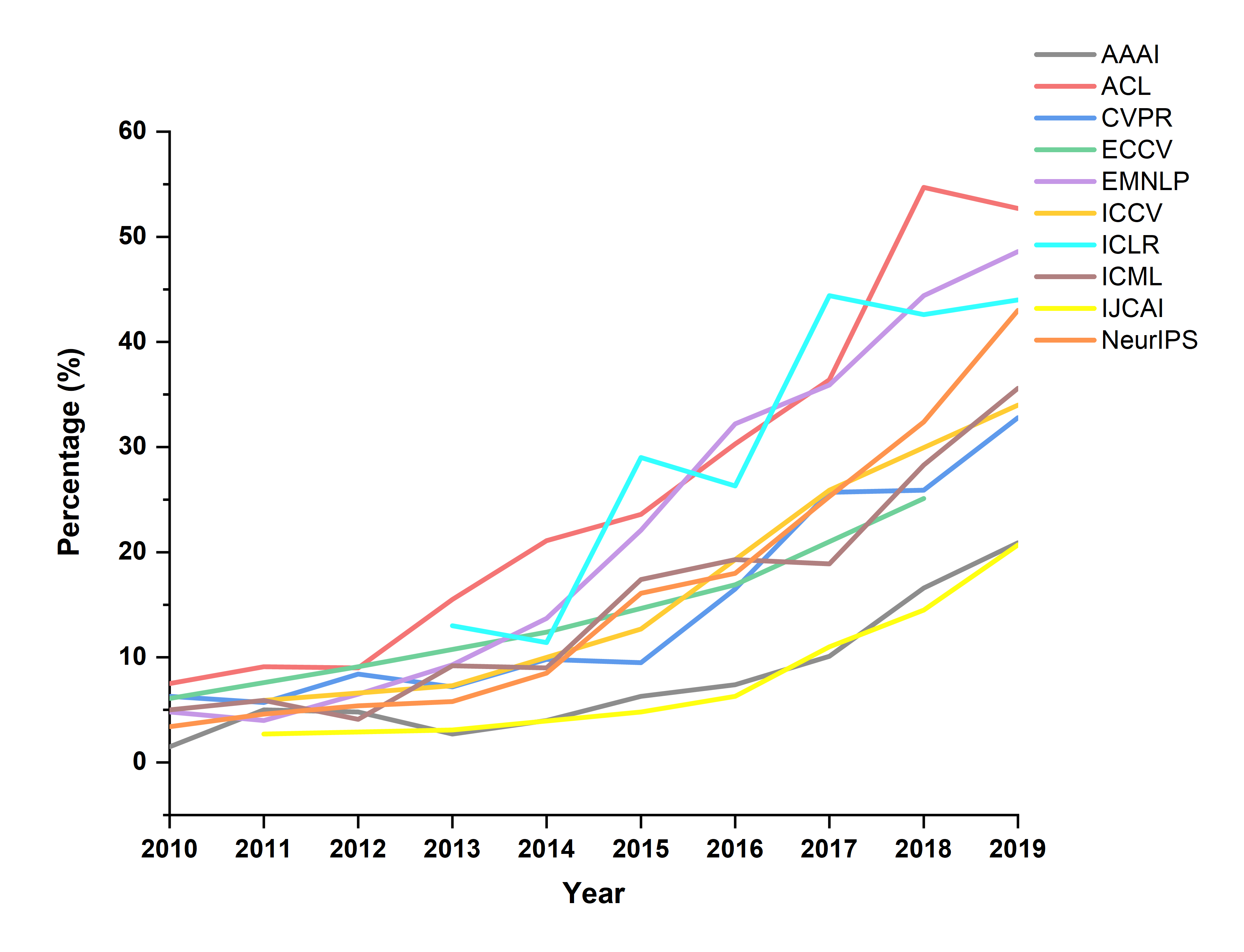}
\caption{Percentage of 10 top AI conference papers with available source code, 2010--2019. For IJCAI from 2011 to 2015 and biennial ECCV and ICCV, the data points for every other year are connected together.}
\label{fig-open-source-code-rate}
\end{figure}

In total, 20.5\% of papers are identified as papers with available source code.\footnote{Errors in the results generated by our model are manually corrected.} Whereas, only around half of them are indexed as open source code papers by Papers With Code. It shows that the proposed method is effective in mining papers with available source code. Fig.~\ref{fig-open-source-code-rate} shows the percentage of papers with available source code of the 10 top AI conferences selected published from 2010 to 2019. The full collection of papers with available source code and their source code repository URLs is presented on our website.\footnote{All resources of this paper can be found at http://www.linjialiang.net/publications/ai-papers-code-analysis .} We will keep the collection project maintained for correction and update.

As shown in Fig.~\ref{fig-open-source-code-rate}, the available source code rate of papers in all 10 top AI conferences increased from 2010 to 2019. The top three conferences are ICLR, ACL, and EMNLP. Since 2018, the percentage of available source code papers on ACL has exceeded 50\%. EMNLP, ICLR, and NeurIPS in 2019 also register a high available source code rate of over 40\%. NLP conference papers have higher available source code rates, whereas CV conference papers have relatively lower available source code rates. AAAI and IJCAI are found to be at the bottom in the rank of available source code rate.

\begin{figure}[htb]
\centering
\includegraphics[width=0.9\textwidth]{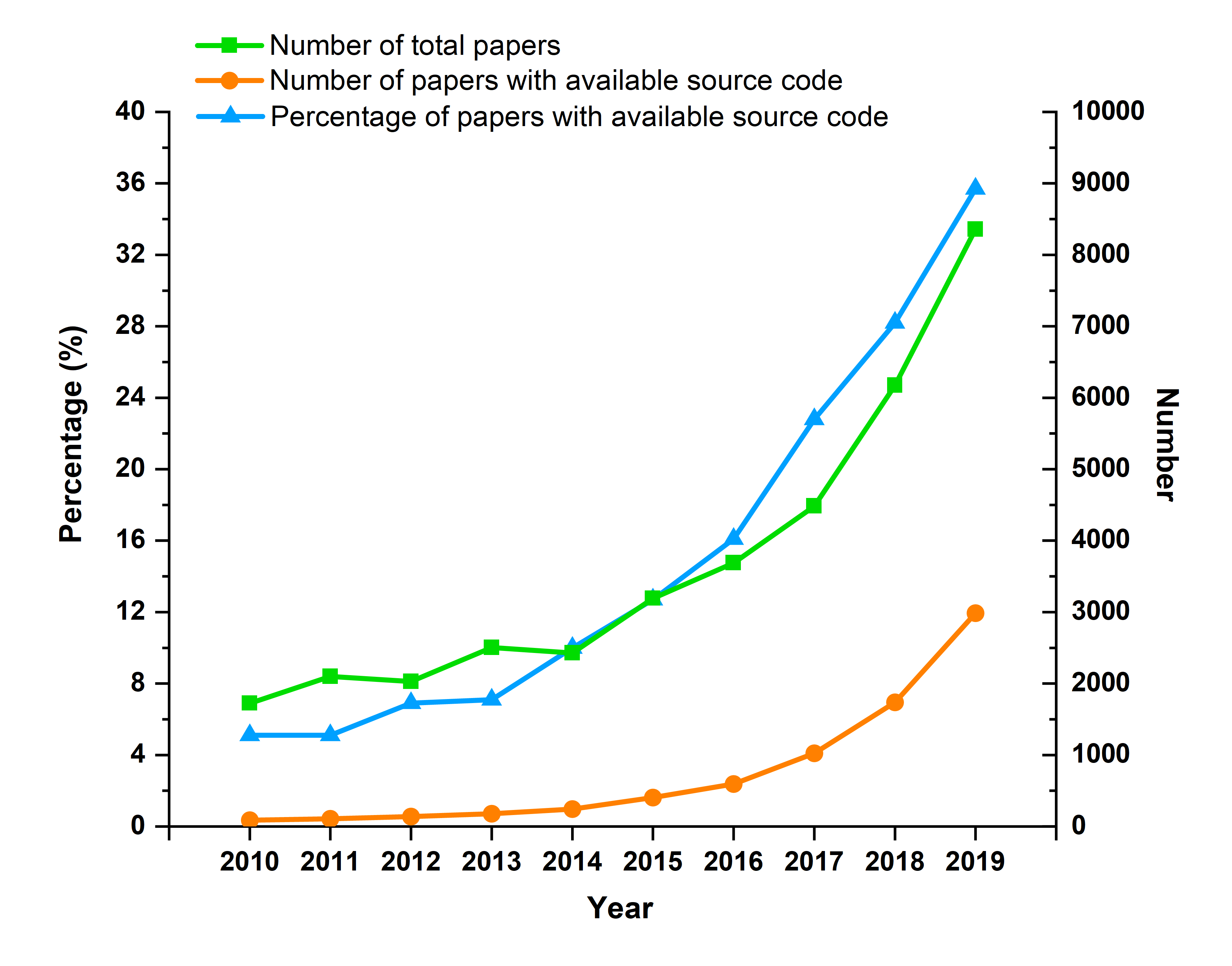}
\caption{Number and percentage of 10 top AI conference papers with available source code, 2010--2019.}
\label{fig-general-open-source-code-trend}
\end{figure}

According to Fig.~\ref{fig-general-open-source-code-trend}, the number of total papers, the number of papers with available source code, and the percentage of papers with available source code have all increased over the last 10 years. In 2019, the number of total papers nearly reached 8,500 and the number of papers with available source code approached approximately 3,000. We can see that the percentage of papers with available source code has grown from about 5\% to around 35\% in just 10 years. Releasing source code with a paper is gathering momentum in top AI conferences.

We check the accessibility of every source code repository URL we collect, and 8.1\% of them are found inaccessible. Some URLs are inaccessible because of the reconstruction of websites. Some are deleted or made private, causing 404 errors on GitHub. Some are accessible URLs but with no code provided.

The source code is released together with the published papers and thus the source code should be an indispensable part of the paper. Removing the source code online is essentially equal to destroying the integrity of the paper. This behavior should be avoided. Authors should do their best to keep the accessibility of the source code repository URLs included in their papers. For the case of website reconstruction, a redirected link should be added to the original URL.

\section{RQ2: What Are the Characteristics of These Source Code?}
\label{sec:rq2}
To answer RQ2, we study platforms that hold these source code, the programming languages they are written in, the level of recognition, the level of participation of other developers, and the key content of their corresponding papers.

\subsection{Platform}

\begin{figure}[htb]
\centering
\includegraphics[width=0.6\textwidth]{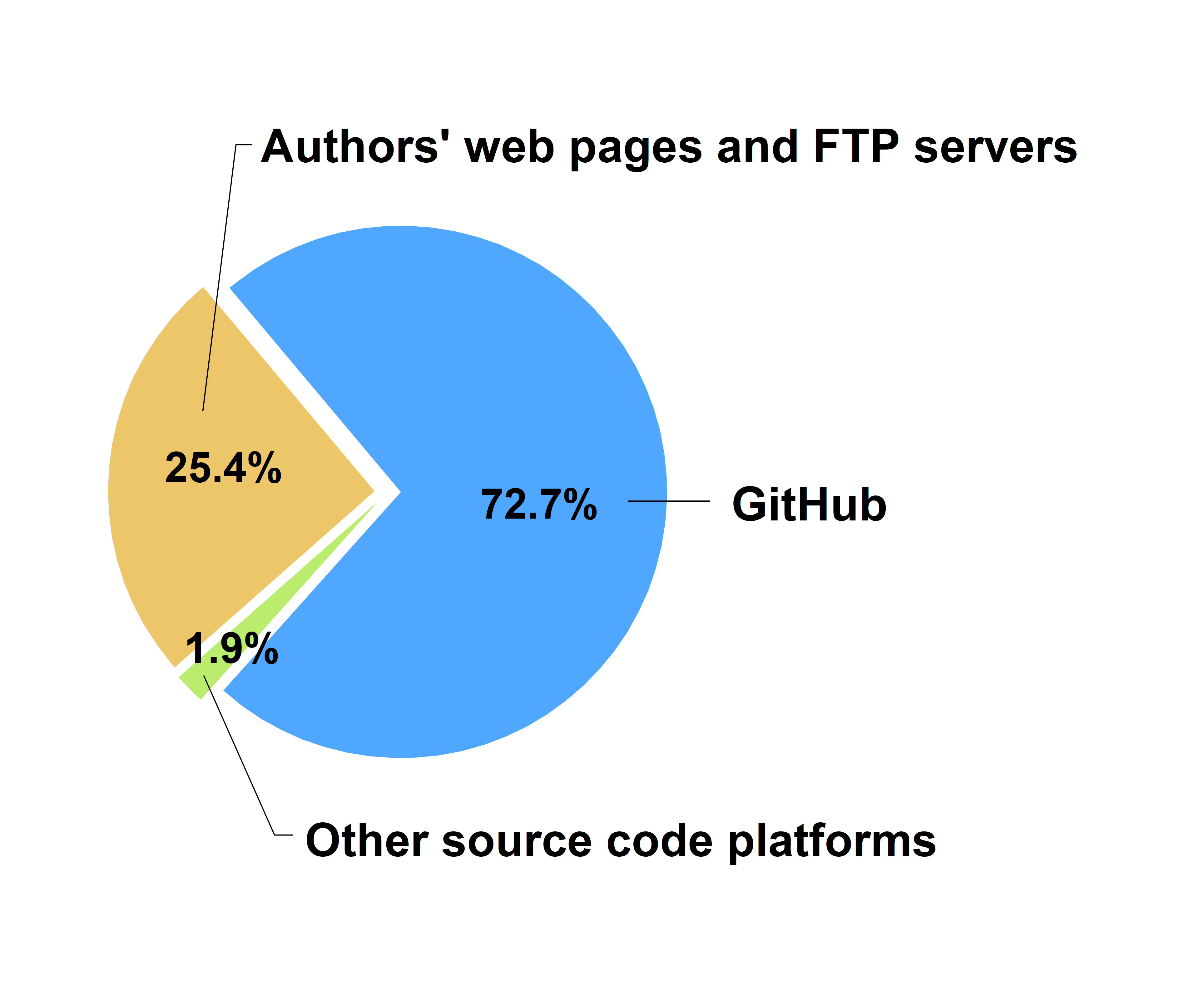}
\caption{Distribution of available source code repositories of papers published in 10 top AI conferences, 2010--2019. The amount of FTP servers is extremely tiny.}
\label{fig-domain-distribution}
\end{figure}

We first conduct a statistical analysis on the domains of source code repository URLs collected in Section~\ref{sec:rq1} to find out what platforms host them. See Fig.~\ref{fig-domain-distribution} for detailed information.

As shown in Fig.~\ref{fig-domain-distribution}, most of the authors choose GitHub\footnote{https://github.com/} to host their source code. Only a very small fraction of them turn to other source code platforms, e.g. Bitbucket,\footnote{https://bitbucket.org/} Google Code,\footnote{https://code.google.com/} CodaLab,\footnote{https://codalab.org/} GitLab\footnote{https://gitlab.com/} and SourceForge.\footnote{https://sourceforge.net/} In addition, nearly one-third of the authors release their source code on their own web pages\footnote{Some are web pages with descriptions, resources, GitHub repository links, etc. These URLs are considered as this kind.} and FTP servers.

GitHub has become the de facto source code platform for the AI community. It allows users to access the repository metadata through a universal API.\footnote{https://docs.github.com/en/rest . The API data were collected on Jan 25, 2022.} Thus we use GitHub-held repositories as representative samples in our study of programming languages, the level of recognition, and the level of participation for RQ2.

\subsection{Programming language}

We then set out to study the programming languages these repositories use. Among them, Python is the most popular. It is written in a highly concise and quick manner. There are also a large number of machine learning and analytical packages for Python users. With the growing importance of deep learning, it is crucial to enable GPU in computation. In this respect, Python is supported by numerous frameworks, e.g. Theano \cite{bergstra-theano-2010}, Caffe \cite{jia-caffe-2014}, TensorFlow \cite{abadi-tensorFlow-2016} and PyTorch \cite{paszke-pytorch-2019} to accelerate GPU-based deep learning. All these strong sides above turn Python into the first choice for AI programming. Jupyter Notebook\footnote{The metadata we get from GitHub API regards it as a programming language.} is the second most popular one. In fact, most work in Jupyter Notebook is written in Python. The difference between them lies in that Jupyter Notebook fosters more interaction. C++ ranks third. This traditional language has its own advantage in speed and it is often used in speed-sensitive projects. The fourth place is taken by MATLAB, a famous script language for simulation and evaluation of research ideas. See Table \ref{tab-language-distribution} for detailed information.

\begin{table}[htb]
		\caption{Programming languages on GitHub repositories of papers with available GitHub source code of 10 top AI conferences, 2010--2019. Data only contain accessible GitHub repositories. *Repositories without language information in GitHub API are classified as "Others".}
		\label{tab-language-distribution}
		\centering
		\begin{tabular}{lll}
				\hline\noalign{\smallskip}
				\textbf{Programming language}  &  \textbf{Percentage} \\
				\hline\noalign{\smallskip}
				Python   &  66.9\%   \\
				Jupyter Notebook  &  7.8\%  \\
				C++   &  5.9\%  \\
				MATLAB    &  5.8\%  \\
				Others*   &  13.6\%  \\
				\hline\noalign{\smallskip}
		\end{tabular}
\end{table}

\subsection{Recognition and participation}

To measure the recognition and the participation level of these source code, we use their number of stars and forks on GitHub. The number of stars can indicate the quality of GitHub repositories to some extent. Stars are awarded to a repository to show users' recognition of its usefulness, inspiration, enlightenment, etc. The number of forks shows how many users intend to participate in the development and maintenance of a GitHub repository. Users fork repositories to make changes or use them as starting points for their own projects. Overall, the number of stars and forks can be regarded as the public rating for a source code repository, and quantitative indicators of its recognition and participation.

For each top AI conference, we count the total number of stars and forks gained by their papers with available source code on GitHub. A statistical test is then performed on the counting results. We first use the D'Agostino-Pearson test \cite{dagostino-omnibus-1971,pearson-tests-1977} to test all data for normal distribution. All of the results reject the null hypothesis at a 5\% significance level, which means that these samples do not look Gaussian. We then apply the Kruskal-Wallis test \cite{kruskal-use-1952} to all ten groups (one conference per group) together to determine if there are significant differences in the mean ranks. At the 5\% level of significance, the null hypothesis is rejected by the result. This shows that the mean ranks of these groups are not the same. Then we use the Mann-Whitney U test \cite{mann-test-1947} to conduct a pair of comparisons on the number of stars and forks, respectively, for the 10 top AI conferences selected. For the number of stars, AAAI \& EMNLP, ACL \& ICML, ACL \& NeurIPS, ECCV \& ICCV, ECCV \& ICLR, ICCV \& ICLR, and ICML \& NeurIPS fail to reject the null hypothesis at a 5\% significance level. This indicates that these pairs share a similar distribution of data. As for the number of forks, AAAI \& EMNLP, AAAI \& IJCAI, ACL \& ICML, ACL \& NeurIPS, ECCV \& ICLR, EMNLP \& IJCAI, and ICML \& NeurIPS follow the similar distribution. The remaining pairs reject the null hypothesis and thus their data follow different distributions. In addition, many extreme values are found in the data. Taking all the above factors into consideration, the median is chosen to analyze the data. See Table \ref{tab-conference-star-fork-number} for detailed information.

\begin{table}[htb]
	\caption{Median numbers of stars and median numbers of forks on GitHub repositories of papers with available GitHub source code of 10 top AI conferences, 2010--2019. Data only contain accessible GitHub repositories.}
	\label{tab-conference-star-fork-number}
	\centering
	\begin{center}
		\resizebox{\textwidth}{!}{
		\begin{tabular}{cccccccccccc}
			\hline\noalign{\smallskip}
			\textbf{Conference} & \textbf{AAAI} & \textbf{ACL} & \textbf{CVPR} & \textbf{ECCV} & \textbf{EMNLP} & \textbf{ICCV} & \textbf{ICLR} & \textbf{ICML} & \textbf{IJCAI} & \textbf{NeurIPS} & \textbf{All} \\
			\hline\noalign{\smallskip}
				Median number of stars & 17 & 30 & 113.5 & 71.5 & 20 & 63 & 73 & 24 & 11 & 30 & 37 \\
				Median number of forks & 5  & 7  & 26  & 18.5 & 5  & 14 & 18 & 7  & 4  & 8  & 9  \\
			\hline\noalign{\smallskip}
		\end{tabular}
	}
	\end{center}
\end{table}

The Spearman's rho \cite{spearman-proof-1904} between the median number of stars and the median number of forks is 0.976 with a \textit{P} value smaller than 0.01. This indicates a strong correlation between them. Among 10 conferences, CVPR earns the most stars and forks as represented by the median. It should be noted that CVPR also achieves the highest h5-index among all these conferences.\footnote{https://scholar.google.com/citations?view\_op=top\_venues} ICLR and ECCV take the position of the second and the third with the similar median values of stars and forks. On a median basis, IJCAI has the lowest number of stars and forks. Its median number of stars is much lower than that of other conferences. In general, conferences of CV have a higher median value of stars and forks than conferences of NLP. This suggests that, to a certain degree, the CV code is more popular than the NLP code.

The top 10 repositories with the most stars are listed in Table \ref{tab-top-star-repository}.\footnote{Some of the repositories match more than one paper. Because of the limited space, the titles of the papers are not presented in the table.} We can see that most of them are projects developed by IT giants. This to some extent shows that the IT industry carries a big weight in the development of AI.

\begin{table}[htb]
\caption{Top 10 most-starred GitHub repositories of papers with available GitHub source code of 10 top AI conferences, 2010--2019.}
\label{tab-top-star-repository}
\centering
\begin{tabular}{ll}
\hline\noalign{\smallskip}
\textbf{Repository URL} & \textbf{Number of stars} \\
\hline\noalign{\smallskip}

https://github.com/tensorflow/models  &  72,591     \\

https://github.com/facebookresearch/Detectron  &  24,943     \\

https://github.com/google-research/google-research  &  21,627   \\

https://github.com/open-mmlab/mmdetection  &  18,109   \\

https://github.com/tensorflow/magenta  &  17,382   \\

https://github.com/facebookresearch/faiss  &  16,064   \\

https://github.com/pytorch/fairseq  &  15,613   \\

https://github.com/slundberg/shap  &  15,204  \\

https://github.com/rusty1s/pytorch\_geometric  &  13,651  \\

https://github.com/microsoft/LightGBM  &  13,402  \\

\hline\noalign{\smallskip}
\end{tabular}
\end{table}

\subsection{Key content}

\begin{figure*}[htb]
\centering
\includegraphics[width=0.9\textwidth]{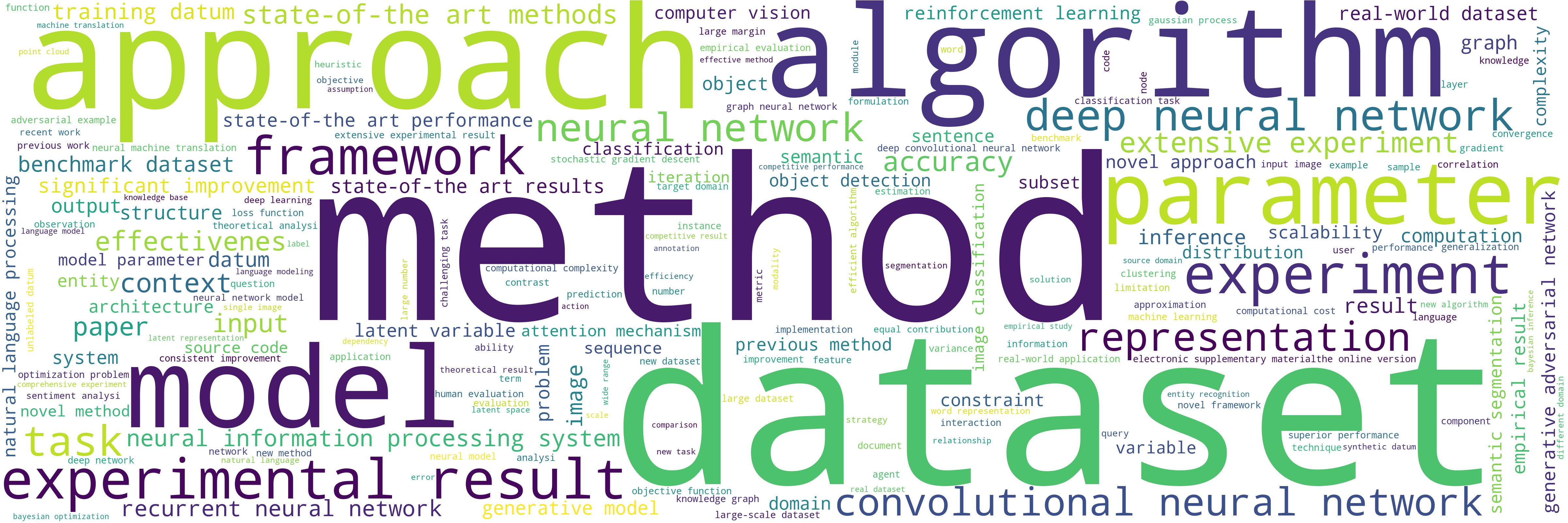}
\caption{Keyphrase word cloud of papers with available source code of 10 Top AI conferences, 2010--2019.}
\label{fig-paper-word-cloud-general}
\end{figure*}

To investigate the key content of the corresponding papers of the source code, we use EmbedRank \cite{smires-simple-2018} to extract the keyphrases from their abstracts. Then we use WordCloud \footnote{https://github.com/amueller/word\_cloud} to generate a word cloud for the keyphrases extracted. See Fig.~\ref{fig-paper-word-cloud-general} for detailed information.

As shown in Fig.~\ref{fig-paper-word-cloud-general}, "method", "dataset", "approach" and "algorithm" are the most common words in the abstracts of top AI conference papers with available source code. It is noteworthy that "dataset" occupies the same prominent position as "method". This demonstrates that datasets play a critical role in AI research. Most AI researchers have their source code released together with the related datasets. It is pleasing to see this behavior becoming prevalent because it works greatly to the benefit of the community.

\section{RQ3: Do the Authors Provide Enough Documentation for Readers to Run the Source Code?}
\label{sec:rq3}

The purposes of releasing source code are many and varied. Some researchers release the code "in order to encourage open research and facilitate future comparisons with the proposed method" \cite{alabort-unifying-2015}. Some researchers are "for the benefit of the community" \cite{booth-3d-2016} or "for the purpose of reproduction and extensions" \cite{qian-improving-2019}. There are also researchers releasing the code "to facilitate future research in this field" \cite{hu-open-2019}. Nowadays, even though top AI conferences do not put authors under obligation to release the source code of their submitted papers, almost all of them require the authors to provide enough technical details for reimplementation. The releasing of source code proves reassuring to the conferences about the quality and thus encourages acceptance of the paper.

However, releasing source code itself does not guarantee a successful reproduction. Authors should provide sufficient documentation to walk their readers through the implementation. Otherwise, releasing source code with no documentation provides little help for reproduction.

Today, the README file is one of the most ubiquitous forms of documentation found in software development. As the name implies, the README file is often the first thing a developer reads when working with a new source code repository. It should provide information like what this repository is and how to use it.

\subsection{Method}

GitHub repositories are documented in README files.\footnote{https://docs.github.com/en/repositories/managing-your-repositorys-settings-and-features/customizing-your-repository/about-readmes} They are written in plain Markdown\footnote{https://docs.github.com/en/get-started/writing-on-github/getting-started-with-writing-and-formatting-on-github/about-writing-and-formatting-on-github} and are converted into HTML to be displayed automatically at the web pages. In our study, we analyze the main README files shown on the repository profile pages of the GitHub repositories.

\begin{figure}[htb]
\centering
\includegraphics[width=0.6\textwidth]{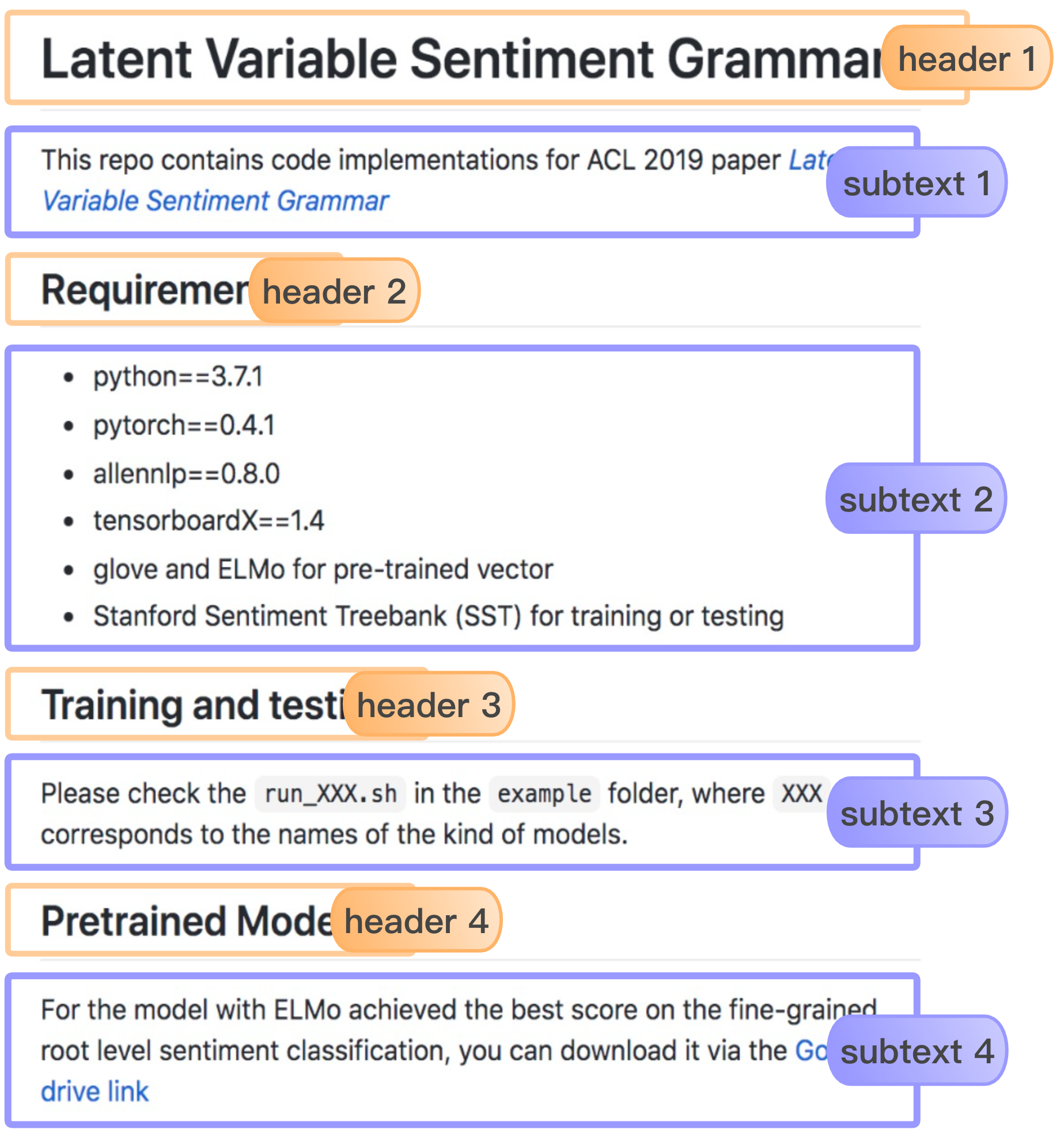}
\caption{Segmentation for a sample README file. Details of the example README file can be found at https://github.com/Ehaschia/bi-tree-lstm-crf .}
\label{fig-readme-structure}
\end{figure}

A README file can be of great length and thus cannot be analyzed with the whole file as one unit. In terms of structure, a README file is composed of several sections and/or subsections, each with corresponding headers. The structure of README file is not unified as it may or may not include subsections. For this reason, we separate a README file by all the headers, regardless of their levels, in other words, we take section headers and subsection headers as the same. A header and the following text (called "subtext") are treated as one unit. Fig.~\ref{fig-readme-structure} illustrates our method of README file segmentation. All the units of a README file are collected into a union set. This union set then reveals the information contained in the README file.

README files of available source code repositories in AI papers usually contain similar contents. By observing and summarizing a huge number of README files, we divide the content of these README files into eight categories. One unit can fall into one or more categories.

\begin{itemize}
\item \textbf{Acknowledgment}
Acknowledgment to other source code, papers, fundings, or any kind of help to this repository. References are considered as Acknowledgment in our study;

\item \textbf{Citation}
Citation information of the corresponding papers of the repository. The contact information is commonly included in the citation information, and it is also classified into this category;

\item \textbf{Installation}
Installation instructions for the repository, including prerequisites, requirements, dependencies, etc.;

\item \textbf{License}
Licenses for using the repository and copyright declarations;

\item \textbf{Others}
Contents that do not belong to any of the other categories;

\item \textbf{Resource}
Related datasets, word embeddings, and other necessary resources to run the repository;

\item \textbf{Technicality}
General description of the repository or the corresponding paper, including a directory of the README file, folder structure, experiment results, trained models, etc.;

\item \textbf{Usage}
Step-by-step instructions on how to use the repository.

\end{itemize}

\begin{figure}[htb]
\centering
\includegraphics[width=0.6\textwidth]{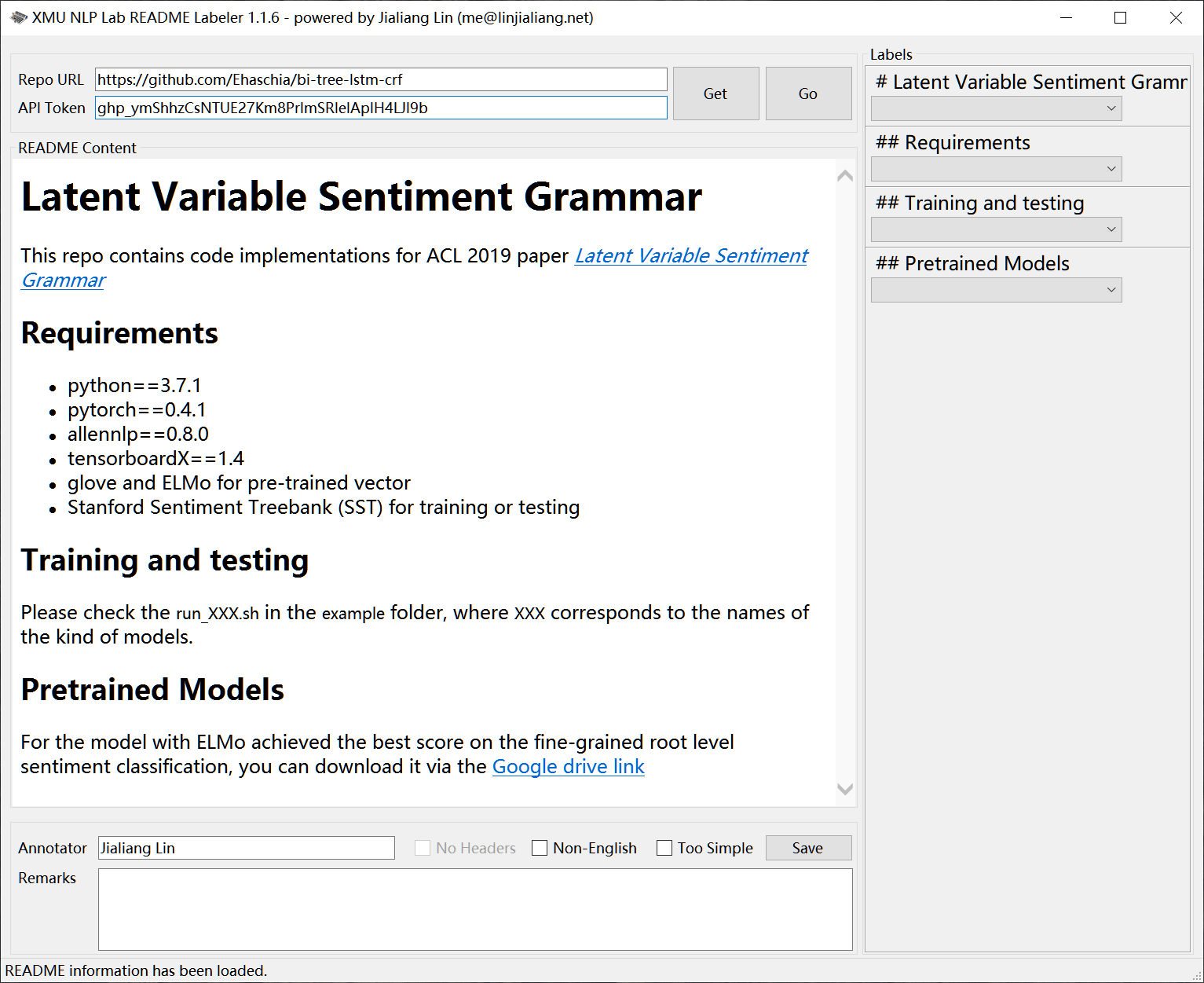}
\caption{The interface of the XMU NLP Lab README Labeler.}
\label{fig-readme-labeler-snapshot}
\end{figure}

We develop the XMU NLP Lab README Labeler to label these README files based on the categories presented above. This tool can load the content of a README file and extract headers as the representatives of units for labeling in a multi-selection manner. There is a "Non-English" option to mark if the content has non-English characters. We also include an option of "Too Simple" to mark those README files with overly insufficient information provided. Fig.~\ref{fig-readme-labeler-snapshot} shows the interface of the XMU NLP Lab README Labeler. Compared to labeling in a purely manual manner, our tool delivers effective and accurate performance on labeling tasks because the chances of errors are reduced with automation. On average, it takes approximately 4.5 minutes for a README file to be labeled by annotators of Prana et al. \cite{prana-categorizing-2019}, whereas, with our tool, it only takes about 2 minutes. This tool is available on our website.

We have 10 annotators, with 4 of them being the authors of this paper, to annotate the headers of the README files using our tool. All the annotators are trained together under the same labeling guideline. When labeling a README file, they all work on it at the same time and are instructed to communicate with each other to ensure consistent labeling. A header can be assigned with one or more labels. Each README file is labeled by two non-author annotators at first. The inter-annotator agreement metric Cohen's kappa \cite{cohen-coefficient-1960} of the first round labeling is 0.685. The disagreements of the first round labeling are resolved by one of the author annotators.

\subsection{Analysis}

In total, we label 5k unduplicated README files. With these labeled data, we create a README dataset named the XMU NLP Lab README Dataset. This README dataset is larger than the README dataset of Prana et al. \cite{prana-categorizing-2019} that is composed of 393 README files. The labeled files and the original README files can be downloaded from our website. Among these README files, 5.8\% of them are labeled "Too simple", which indicates that these files are inadequate in terms of information provided for users to run the source code. This inadequacy speaks to an underlying problem in the open source community. Some authors might choose to release their source code to the public, but they are not fully involved in the opening as they fail to provide enough information to actually run the source code. Full results of the labeling are listed in Table \ref{tab-readme-header-percentage}. A small number of README files do not provide header information and they are excluded from the analysis.

\begin{table}[htb]
\caption{Percentage of README units of different categories. Units can have multiple categories.}
\label{tab-readme-header-percentage}
\centering
\begin{tabular}{lll}
\hline\noalign{\smallskip}
\textbf{Category} &  \textbf{Percentage} \\
\hline\noalign{\smallskip}

Acknowledgment  &  26.4\%   \\

Citation    &  88.2\%   \\

Installation    &  61.9\%  \\

License   &  14.4\%  \\

Others   &  4.1\%  \\

Resource   &  38.5\%  \\

Technicality    &  79.6\%  \\

Usage   &  61.6\%  \\

\hline\noalign{\smallskip}
\end{tabular}
\end{table}

As presented in Table \ref{tab-readme-header-percentage}, nearly 90\% of the authors provide citation information in README files. This to some degree shows that authors think highly of citation impact and make serious efforts to gain more citations. Almost four-fifths of the authors present related technicality in their README files. Less than two-thirds of authors provide installation instructions or usage tutorials of the repositories. Only around one-third of the README files contain these necessary resources, but some developers might have them provided but not presented in the README files. Approximately, one-fourth of the README files hold acknowledgment information and this number mirrors the great importance of previous work in AI research. Merely 14.4\% of the repositories provide clear licensing information in the README files. Without these licensing details, it is difficult for users to legally comply with the license requirements. Though in some cases, the licensing information is provided in a separate file on GitHub but not included in the README files to avoid duplication.

In summary, judging from the README files of the sampled papers with available source code, quite a fair amount of authors fail to provide the necessary documentation for users. Certain missing information, e.g. installation instructions and usage tutorials, is indeed indispensable for running the available source code. Therefore, we suggest that providing this information in the repository should be made standard practice. In addition, it can be considered a criterion to evaluate reproducibility.

\subsection{Auto-labeling}

Hand labeling is accurate, but this accuracy is gained at the expense of considerable labor and time. It is not an easy task to manually label the already existing README files of papers with available source code published in all top AI conferences. What adds to the difficulty is the fact that there is a massive amount of new papers being published every year. To address this issue, we use a multi-label classification model to label README files automatically.

Our idea is developed from Jeblick \cite{jeblick-two-2020} and Cerliani \cite{cerliani-siamese-2020}. After we finished our experiment independently, we found that Jaunjale \cite{jaunjale-scibert-2020} had proposed a similar idea. The architecture of the classification model is generally the same as the model described in Section \ref{subsec:method}, with two main changes. First, the single input becomes a dual input. In the labeling method mentioned above, we take a header and its subtext as one unit. In terms of length, a header is much shorter than its subtext. But because a header outlines the content of the subtext, the header plays a greater role in the final label. Therefore, a header and its subtext should not be simply concatenated and encoded together. Based on this, we use two independent SciBERT to encode the header and the subtext separately as the input. Second, the output has changed from single-label to multi-label. While the original model needs only one label to indicate if the input sentence provides the source code information of the paper itself, the model here requires eight labels to represent whether the input unit belongs to the corresponding categories or not. We use binary cross-entropy as the loss function and sigmoid as the last layer's activation function. Our README dataset is randomly divided into a training set, a validation set, and a test set at a ratio of 8:1:1 and used for model training and evaluation.

Prana et al. \cite{prana-categorizing-2019} also train models to label GitHub README files but with traditional machine learning methods. We use its source code to evaluate our README dataset with the same training set, validation set, and test set as a baseline. Our model outperforms theirs in both accuracy and weighted F1-score. See Table \ref{tab-performance-comparison} for detailed information.

\begin{table}[htb]
	\caption{Comparisons of accuracy and weighted F1-score of README labeling methods.}
	\label{tab-performance-comparison}
	\centering
	\begin{tabular}{ccc}
	\hline\noalign{\smallskip}
					    	   		&  \textbf{Accuracy}   	&    \textbf{Weighted F1-score}		 \\
    \hline\noalign{\smallskip}
		Proposed method    			&   \textbf{0.779}         &      \textbf{0.850}         		 \\
		Compared method \cite{prana-categorizing-2019}  &  0.620           &  0.671         		 \\
	\hline\noalign{\smallskip}
	\end{tabular}
\end{table}

\section{Pipeline of the Whole Analysis}
\label{sec:whole-pipeline}

\begin{figure}[htb]
\centering
\includegraphics[width=0.8\textwidth]{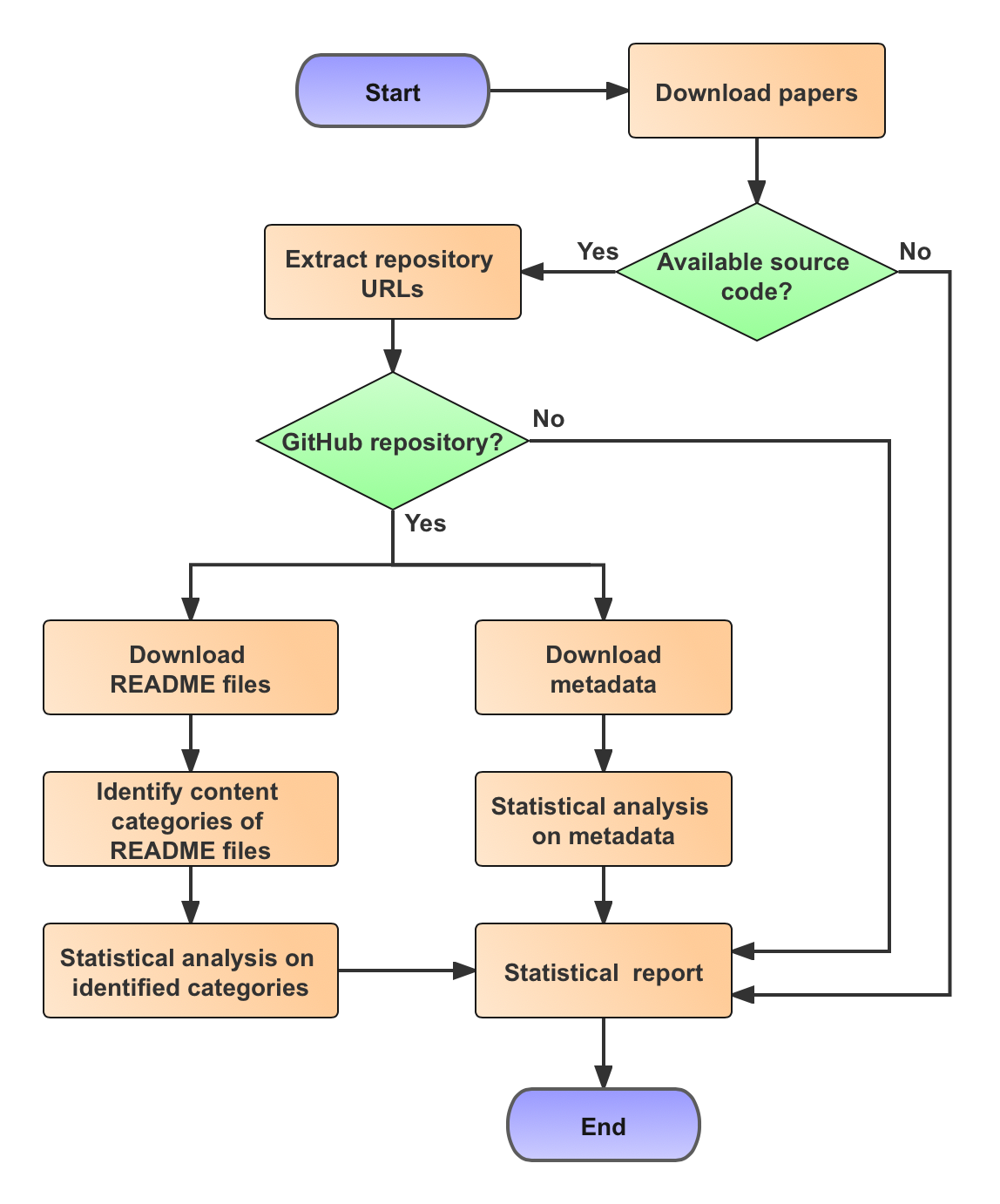}
\caption{The pipeline of analysis of papers with available source code of an AI conference.}
\label{fig-flow-analyze}
\end{figure}

The whole analysis pipeline is shown in Fig.~\ref{fig-flow-analyze}. The innovation in our automatic analysis mainly lies in two aspects. In the first place, we design an automatic SciBERT-based classification model to identify whether or not the paper has the available source code. We feed corresponding sentences of URLs into our model and then get the result of classification. Our model also extracts the source code repository URL of each paper's available source code during this process. In the second place, we use a dual-SciBERT model to identify the content categories of README files of GitHub source code repositories. After doing that, we can evaluate the actual reproducibility based on the sufficiency of the README files for source code running.

\section{Related Work}
\label{sec:related-work}

The work presented in this paper is related to research in the following four areas: statistical analysis on papers with available source code, mining of GitHub data, research on text classification, and studies of software documents.

\subsection{Statistical analysis on papers with available source code}
Many contributions have been made to the collection and analysis of papers with available source code. Papers With Code links source code repositories to AI papers and categorizes them according to their research topics with benchmarks provided. MLNLP World manually collects papers with available source code that are published in the top NLP and general AI conferences.\footnote{https://github.com/MLNLP-World/Top-AI-Conferences-Paper-with-Code} Vandewalle \cite{vandewalle-code-2012,vandewalle-code-2019} analyzes papers with available source code published in \textit{IEEE Transactions on Image Processing} 2004--2006 and 2017. He finds that there is a 10\% increase in the number of such papers over the last decade and that on average, these papers gain more citations than those without. Russell et al. \cite{russell-large-2018} collect 1,720 GitHub repositories of bioinformatics papers and conducts an analysis on their source code, metadata, and profile of papers. Bonneel et al. \cite{bonneel-code-2020} manually assess the replicability of papers with available source code published in SIGGRAPH 2014, 2016, and 2018. Färber \cite{farber-analyzing-2020} is the first to conduct a thorough analysis on 2,955 GitHub repositories based on the Microsoft Academic Graph (MAG) \cite{sinha-overview-2015}.

\subsection{Mining of GitHub data}
A large number of researchers have studied GitHub, the world's largest source code platform, from various aspects. Ray et al. \cite{ray-large-2014} conduct a large-scale statistical analysis on programming languages with GitHub data. Cosentino et al. \cite{cosentino-findings-2016} summarize 93 papers that have their available source code hosted by GitHub and performs a meta-analysis on their empirical methods, datasets, and limitations. In order to reduce the risk of using unmaintained open source projects and to attract collaborative efforts in the maintenance of these projects, Coelho et al. \cite{coelho-identifying-2018} train models to identify unmaintained or rarely maintained projects on GitHub based on their commits, forks, and feedback on issues. Kim \& Lee \cite{kim-understanding-2021} conduct comprehensive research on the user working habits of the activities between GitHub commits and Stack Overflow posts. Wattanakriengkrai et al. \cite{wattanakriengkrai-github-2022} perform a comprehensive statistical analysis of scholar papers in GitHub repositories created in years between 2014 and 2018 to study the public accessibility, traceability, and evolution of links to scholar papers.

\subsection{Research on text classification}
Text classification is an essential NLP technology. Recent years have seen the rise of deep learning methods for text classification. Liu et al. \cite{liu-recurrent-2016} make use of recurrent neural network to learn text representation and their work achieves promising results. Conneau et al. \cite{conneau-very-2017} construct a new architecture VD-CNN with up to 29 convolutional layers for text classification. This method delivers state-of-the-art performance on several public text classification tasks. Yao et al. \cite{yao-graph-2019} propose a Text Graph Convolutional Network. This simple network of two layers outperforms numerous state-of-the-art methods on the classification tasks of several benchmark datasets. Wang et al. \cite{wang-novel-2021} introduce ML-Reasoner, a reasoning-based multi-label classification model. This model can make use of information from different labels to overcome the problem of label-order sensitivity. Minaee et al. \cite{minaee-deep-2021} conduct a comprehensive review of more than 40 datasets used for text classification and over 150 deep learning-based text classification models. In the review, comparisons are made between these models on their contributions and advantages. In addition, an analysis is performed on the well-known benchmarks of these models.

\subsection{Study of software documents}
Software documentation is an important component of software that provides users with great help in deploying, using, and maintaining of the software. Abebe et al. \cite{abebe-empirical-2016} study contained information in release notes of software to give developers empirically-supported suggestions in the writing of these notes. Hassan \& Wang \cite{hassan-mining-2017} use README files as a source for automatic extraction of software build commands. Zhang et al. \cite{zhang-detecting-2017} propose a method to detect similar repositories on GitHub based on similar contents of the README files. Li et al. \cite{li-to-2018} highlight the importance of negative caveats of API documentation so that programmers can avoid the improper use of APIs. Hassan et al. \cite{hassan-usability-2020} introduce a usability requirements extraction approach based on a usability keywords repository created by the usability category of ISO 9126 and ISO 25010. Nath \& Roy \cite{nath-automatically-2021} propose a release notes generation method by employing text summarization technologies. They then classify the generated notes into 6 categories for friendly reading.

\section{Future Work and Conclusion}
\label{sec:future-conclusion}

We will continue our study to other aspects of the original 10 top AI conferences selected. Meanwhile, we are going to extend our project to include papers on other conferences or even journals. Our vision is to undertake a panoramic study on papers with available source code of all top AI conferences and top AI journals.

In this paper, we proposed an automatic method to identify AI conference papers with available source code and extract URLs of their source code repositories. This method can also be applied to all other scientific papers. Our method reached an accuracy of 0.939. With this method, 20.5\% out of 36,679 regular papers published in 10 top AI conferences from 2010 to 2019 were identified as papers with available source code. Their source code repository URLs were extracted and provided on our website. Through analysis of these papers, we found that there was a growing trend in releasing source code along with papers over the last decade. However, 8.1\% of the source code repositories were observed to be inaccessible. This practice of disabling the source code should be discouraged. Of these source code repositories on GitHub, Python was found to be the most popular programming language. As measured by the median, repositories of papers published in CVPR gained the most stars and forks on GitHub and this was a reflection of its high quality. Further, we constructed and release the XMU NLP Lab README Dataset of 5k manually labeled README files from the GitHub repositories of the papers with available source code identified from the 10 top AI conferences. With analysis conducted on this dataset, we discovered that 88.2\% of the authors provided citation information in their README files and over one-third of them did not include installation instructions or usage tutorials for users. Lastly, we used a dual-SciBERT multi-label classification method to label the categories of README files automatically. Our methods serve as an automatic solution to the statistical analysis of AI conference papers with available source code. This solution helps us achieve a comprehensive understanding of the current situation of source code in the AI community. With this, we aim to prompt more developments in the open source phenomenon.

\section*{Acknowledgments}
\label{sec:ack}
This work is partly funded by the 13th Five-Year Plan project Artificial Intelligence and Language of State Language Commission of China (Grant No. WT135-38). We appreciated Yuna Chen, Wen Yuan, Zhiyang Zhou, Shuangtao Li, Jiaxin Song, and Zhaohong Lai for their help in data labeling. We also thank two anonymous reviewers for their useful comments. Special and heartfelt gratitude goes to the first author's wife Fenmei Zhou, for her understanding and love. Her unwavering support and continuous encouragement enable this research to be possible.

\bibliographystyle{ws-ijseke}
\bibliography{ws-ijseke}

\end{document}